# MxDiffusion: A Physics-Aware Maxwell's Law-Guided Diffusion Model Strategy for Inverse Photonic Metasurface Design

*Sujoy Mondal[1], Taehyuk Park[1], Sudipta Biswas[2], Alan X. Wang[2], Wenshan Cai[1,3]\**

[1] School of Electrical and Computer Engineering, Georgia Institute of Technology, Atlanta, Georgia 30332, United States

[2] Department of Electrical and Computer Engineering, Baylor University, Waco, Texas 76798, United States

[3] School of Materials Science and Engineering, Georgia Institute of Technology, Atlanta, Georgia 30332, United States

*Correspondence to: wcai@gatech.edu

**Abstract.** We introduce MxDiffusion, a hybrid physics- and data-driven diffusion-based framework that enables efficient and highly accurate generation of photonic structures from target optical properties. The improved accuracy is achieved through a two-stage generation strategy, in which the first diffusion model is explicitly trained with Maxwell's equation–based loss to embed physical insight directly into the inverse design process, while the second model maps the physically consistent intermediate representation to the final structural geometry with significantly higher fidelity than solely data-driven approaches. The performance of MxDiffusion is validated on two representative applications: gold nanostructures patterned on a silica substrate and a highly tunable bandpass filter based on phase change material. In both cases, the proposed framework consistently outperforms a conventional data-driven diffusion model benchmark, particularly for out-of-training-distribution design targets and highly constrained resonance conditions. These results demonstrate the efficacy and superiority of MxDiffusion as a general physics-guided inverse design paradigm.

**Introduction.** Machine Learning (ML)-based photonic inverse design refers to the use of machine learning models to determine structural parameters that satisfy a prescribed optical response. Due to the time-consuming and inefficient nature of traditional trial-and-error approaches for designing photonic structures, machine learning–based techniques have emerged as powerful tools for exploring complex design objectives that are difficult or even impractical to achieve using trial-and-error methods.[1] ML-based inverse design of photonic structures has attracted significant attention over the past several years [2-16] and have been adopted across many domains of optics and photonics, including nonlinear optics,[17, 18] quantum photonics,[19-21] imaging systems,[22-24] photonic crystals,[25] metasurface design,[26] to name a few.

While deep neural networks (DNNs), such as fully connected networks (FCNs) and convolutional neural networks (CNNs), are well suited for forward prediction tasks,[27, 28] for example, predicting the transmission spectrum a metasurface, inverse design problems typically require generative models, such as variational autoencoders (VAEs) and generative adversarial networks (GANs), to produce new structural configurations that satisfy specified optical response targets. A variational autoencoder (VAE) is a deep generative model composed of an encoder and a decoder, where the encoder compresses input data into a low-dimensional latent representation and the decoder reconstructs the original data by minimizing reconstruction error during training. Once trained, the decoder can generate new samples from random latent vectors. In photonic inverse design, VAEs have been employed as pattern generators in combination with forward prediction networks or further refined using evolutionary strategies.[17, 29, 30] However, VAEs suffer from limited conditional generative capability, as they primarily generate samples through random latent-space sampling, which restricts precise control over the generated patterns to satisfy the complex design objectives. GANs have become the most widely used generative framework for photonic inverse design.[31-34] A GAN consists of a generator that produces synthetic samples and a discriminator that distinguishes between real and generated data. Through adversarial training, the generator learns to produce outputs that increasingly resemble the true data distribution. Despite their success, GANs rely solely on adversarial feedback without an explicit likelihood formulation or physical constraint, which sometimes leads to unstable training dynamics, mode collapse, and limited diversity in generated designs.[35]

To address these limitations, physics-based techniques such as adjoint optimization and physics-informed loss functions have been integrated with machine learning models to improve inverse design performance.[35-39] More recently, diffusion models have been introduced as a robust alternative for inverse design, as they decompose the generation process into a sequence of noise-prediction steps governed by a Markov chain.[40] Diffusion models have demonstrated superior performance over VAEs and GANs across a wide range of applications,[41-48] including photonic inverse design using conditional inputs such as S-parameters and far-field spatial power distributions. Positional sinusoidal encoding is used as a conditional data for training diffusion model as well.[49] Recent efforts have also incorporated adjoint optimization into diffusion models to inject physical knowledge during sampling;[50] however, this approach requires real-time electromagnetic simulations to compute adjoint gradients, resulting in substantial computational overhead.

In this work, we introduce a physics-aware MxDiffusion framework, which directly integrates Maxwell's equations into the training of a diffusion model, providing physical insight intrinsically within the learning framework. This approach significantly enhances the capability of conventional data-driven diffusion models, particularly in optimizing critical boundary regions of photonic structures to meet target design specifications. Importantly, the proposed method does not require real-time simulations during sampling, and both the training and sampling costs remain comparable to those of traditional data-driven diffusion models. The efficacy of our MxDiffusion framework is illustrated through two sets of design examples. First, we consider a periodic gold nanostructure with fixed thickness and arbitrary geometrical shape patterned on glass substrates, where user-defined transmission spectra serve as the design targets. In this case, the MxDiffusion framework consistently outperforms the data-driven diffusion baseline, particularly when generating structure patterns to satisfy out-of-training-distribution target spectra. In the second example, we investigate the inverse design of a highly spectra tunable bandpass filter using a phase-change material. The objective of this design task is to maximize the spectral separation between the bandpass filter peaks corresponding to the two phases of the phase-change material. The MxDiffusion framework successfully generates structures with spectral tunability that significantly exceeds that of the best design in the training dataset, whereas the solely data-driven diffusion model barely surpasses the dataset's optimal performance. These results clearly demonstrate the fidelity of the proposed framework to transcend the limitations of data-only diffusion model frameworks.

**Diffusion Model**

A diffusion model operates through two main phases: forward diffusion and reverse diffusion,[51-54] as shown in figure S1. During the forward diffusion process, clean images from the dataset are progressively corrupted by adding Gaussian noise over $T$ discrete timesteps until the images become nearly pure Gaussian noise. At the initial stage, the clean image is denoted by $x_0$. At the first timestep ($t = 1$), Gaussian noise $\varepsilon_1 \sim \mathcal{N}(0, I)$ is added to the image according to

$$x_1 = \sqrt{\alpha_1} x_0 + \sqrt{1 - \alpha_1}\, \varepsilon_1 = \sqrt{\alpha_1} x_0 + \sqrt{\beta_1}\, \varepsilon_1 \tag{1}$$

where the definitions of $\alpha_1$ and $\beta_1$ are provided below. Similarly, at an arbitrary timestep $t$, the noisy image $x_t$ is obtained from the previous image $x_{t-1}$ by

$$x_t = \sqrt{\alpha_t}x_{t-1} + \sqrt{1-\alpha_t}\varepsilon_t = \sqrt{\alpha_t}x_{t-1} + \sqrt{\beta_t}\varepsilon_t \tag{2}$$

Here, $\{\beta_1, \beta_2, \ldots, \beta_T\}$ is a sequence of small positive values satisfying $0 < \beta_t \ll 1$, and $\alpha_t = 1 - \beta_t$. The parameter $\beta_t$ controls the amount of new noise added at timestep $t$, while $\alpha_t$ determines how much of the original signal is retained at that step. It can be shown mathematically (see Supporting Information) that the noisy image at timestep $t$ can be directly expressed in terms of the original clean image $x_0$ as

$$x_t = \sqrt{\bar{\alpha}_t}x_0 + \sqrt{1-\bar{\alpha}_t}\varepsilon \tag{3}$$

where $\bar{\alpha}_t = \prod_{s=1}^{t}\alpha_s$

In this expression, $\sqrt{\bar{\alpha}_t}x_0$ represents the remaining clean signal, while $\sqrt{1-\bar{\alpha}_t}\varepsilon$ corresponds to the accumulated noise after $t$ steps. Typically, the total number of timesteps $T$ is chosen in the range of 500–1000, ensuring that only a small amount of noise is added at each step. By the final timestep ($t = T$), the image is transformed into nearly pure Gaussian noise. Since the forward diffusion process is entirely defined by fixed equations with predefined $\alpha_t$ and $\beta_t$, it does not involve any learnable parameters or neural network training.

The reverse diffusion process is responsible for generating new images by progressively removing noise from a noisy image, as illustrated in Figure S1. This process starts with a highly noisy image and reconstructs a clean sample through a sequence of denoising steps. The reverse update at timestep $t$ is given by

$$x_{t-1} = \frac{1}{\sqrt{\alpha_t}}\left(x_t - \frac{\beta_t}{\sqrt{1-\bar{\alpha}_t}}\varepsilon_\theta(x_t, t)\right) + \sigma_t z \tag{4}$$

This equation describes how to obtain a less noisy image $x_{t-1}$ from the current noisy image $x_t$. Here, $\varepsilon_\theta(x_t, t)$ is the noise predicted by the neural network model. While noise is added explicitly during the forward diffusion process using Eq. 3, the reverse process relies on a U-Net–based neural network, as shown in Figure S2, to predict this noise. $\sigma_t z$, the stochasticity term, where $z \sim \mathcal{N}(0, I)$, introduces controlled randomness into the reverse process. This term enables proper sampling from the learned distribution and prevents the model from collapsing to a single deterministic reconstruction.

As shown in Figure S2, during training, the U-Net neural network learns to predict the noise $\varepsilon_\theta$ given the noisy image $x_t$, the diffusion timestep $t$, and additional conditional inputs such as the transmission spectrum or the electric field corresponding to the original image $x_0$. Reverse sampling is performed over $T$ timesteps. The process begins with a randomly sampled noise image $x_T$. At timestep $T$, the trained model predicts the noise using the conditional input and timestep information, and Eq. 4 is applied to obtain $x_{T-1}$. This procedure is repeated sequentially for timesteps $T-1, T-2, \ldots, 1$ each time using the same conditional input, until the final clean image $x_0$ is obtained. The conditional input therefore guides the reverse diffusion process, ensuring that the generated image satisfies the desired physical or spectral constraints. Other generative models typically attempt to generate images from target inputs in a single step, whereas diffusion models perform generation through $T$ sequential timesteps. By decomposing image generation into a series of small denoising tasks, diffusion models transform a complex generation problem into multiple simpler noise-prediction problems, leading to more stable training and improved sample quality.

**MxDiffusion Framework**

The key innovation of this work is the explicit integration of frequency-domain Maxwell's curl–curl wave equation (Eq. 6) into a diffusion-based U-Net framework for inverse nanophotonic design. In most existing machine-learning approaches for inverse design of nanophotonic structures, the problem is formulated as a direct mapping between a design objective and a structural geometry. The input to the neural network is usually a target optical response (such as a transmission or reflection spectrum), and the output is the corresponding structural design intended to achieve that response. However, this inverse mapping is inherently non-unique: multiple distinct geometries can produce nearly identical optical spectra. As the complexity of the target response increases, learning this one-to-many inverse mapping becomes increasingly difficult for neural networks, often leading to unstable training, mode collapse, or physically implausible designs.

To address these challenges, this work proposes a fundamentally different inverse-design strategy, defined as MxDiffusion framework, that incorporates Maxwell's equations directly into the training loop through a physics-based loss function. The overall architecture of the proposed physics-integrated diffusion framework for the inverse design of photonic structures is illustrated in Figure 1. Instead of attempting to map spectra directly to structural geometries in a single step, the proposed framework introduces the electric field distribution as an intermediate physical representation. In the first stage of the framework, a diffusion model is trained to predict the two-dimensional electric field distribution corresponding to a given target optical response (Figure 1a-1c). In addition to the standard Denoising Diffusion Probabilistic Models (DDPM) noise-prediction (data-driven) loss, a Maxwell's equation-based residual loss is incorporated to enforce compliance with Maxwell's equations. This physics-based constraint ensures that the predicted electric fields are physically realizable and eliminates spurious or non-physical field patterns that might otherwise satisfy the data-driven loss alone. This step constrains the inverse problem and significantly reduces the space for admissible solutions. Moreover, electric field patterns tend to exhibit smoother spatial variations, particularly near material boundaries, compared to binary or sharply defined structural geometries. This smoother behavior further simplifies the learning task. In the second stage, a separate diffusion model is employed to generate the final structural geometry (Figure 1d and 1e). In this step, both the target transmission spectrum and the predicted electric field are used as conditional inputs. The electric field distribution is strongly correlated with the underlying geometry, making it a comparatively straightforward task for the diffusion model to accurately predict electric fields than to directly infer the final structure in a single step.

For each training batch, the data-driven loss follows the standard DDPM formulation. Let the original two-dimensional electric field be denoted by $E_0$. During training, a diffusion timestep $t \in [0, T]$ is randomly sampled. Using Eq. (3), a noisy version of the electric field $E_t$ is generated by adding Gaussian noise $\varepsilon \sim \mathcal{N}(0, I)$. The diffusion model then predicts the noise $\hat{\varepsilon}$ conditioned on the noisy field $E_t$, the timestep $t$, and the spectral conditional input. The data-driven loss is defined as the mean squared error between the true noise and the predicted noise:

$$L_{\text{DDPM}} = \| \varepsilon - \hat{\varepsilon} \|^2 \tag{5}$$

This loss encourages accurate noise prediction across all diffusion timesteps. In addition to the DDPM loss, a physics-based constraint is imposed through Maxwell's loss. The frequency-domain Maxwell's curl–curl wave equation is defined as

$$\nabla \times (\mu_0^{-1} \nabla \times E) - \omega^2 \varepsilon_r E = 0 \tag{6}$$

Where $\mu_0$ is the magnetic permeability of free space, $\varepsilon_r$ denotes the material permittivity, and $E$ is the electric field. From Eq. 5, Maxwell's loss can be defined as

$$L_{\text{Maxwell}}^{(E)} = \frac{1}{N} \sum_{n=1}^{N} \| \nabla \times (\mu_0^{-1} \nabla \times \hat{E}^{(n)}) - \omega^2 \varepsilon_r \hat{E}^{(n)} \| \tag{7}$$

where $\hat{E}^{(n)}$ is the predicted electric field for the $n$-th sample in the batch, $\omega$ is the angular frequency, and $N$ is the batch size. Physically realizable electric fields satisfy Eq. 6 exactly; therefore, the Maxwell's loss ideally evaluates to zero for realistic electromagnetic fields.

However, the diffusion U-Net architecture predicts noise rather than the electric field directly. Consequently, Maxwell's loss cannot be applied directly to the network output. To enable the incorporation of Maxwell's loss, Eq. (3) is rearranged to express the original electric field in terms of the noisy field and the noise:

$$E_0 = \frac{E_t - \sqrt{1-\bar{\alpha}_t}\,\varepsilon}{\sqrt{\bar{\alpha}_t}} \tag{8}$$

This relation implies that if the noisy electric field $E_t$ and the noise are known, the original field $E_0$ can be reconstructed. During training, the original electric field $E_0$ and timestep $t$ are first used to generate the noisy field $E_t$ via Eq. (3). The noisy field $E_t$, together with the spectral conditional input and timestep $t$, is then passed through the diffusion model to predict the noise $\hat{\varepsilon}$. Substituting the predicted noise $\hat{\varepsilon}$ into Eq. (8) yields a reconstructed electric field $\hat{E}_0$. Since $\hat{\varepsilon}$ approximates the true noise, $\hat{E}_0$ deviates from the ground-truth field. The Maxwell loss in Eq. (7) is then evaluated on $\hat{E}_0$ and backpropagated through the network, enforcing physical consistency during training.

For large timestep values, the predicted noise becomes increasingly uncertain, leading to large reconstruction errors and correspondingly high Maxwell's loss. Therefore, Maxwell's loss is applied only for small timestep values, where the reconstructed electric fields remain physically meaningful. In practice, the model is trained in parallel using uniformly sampled timesteps $t \in [0, T]$ for DDPM loss and restricted timesteps $t < 100$ for Maxwell's loss. This combined training strategy effectively balances data fidelity with physical realism. The complete training procedure is summarized in Algorithm 1.

| | | Algorithm 1: Physics-Guided Diffusion Model Training | |
|---|---|---|---|
| | Step | Description | Explanation |
| | | **Repeat** | |
| Normal DDPM Training Stage | 1 | $x_0 \sim q(x_0)$ | Sample clean data |
| | 2 | $t \sim \text{Uniform}(1, \ldots, T)$ | Sample timestep |
| | 3 | $\epsilon \sim \mathcal{N}(0, I)$ | Sample Gaussian noise |
| | 4 | $x_t = \sqrt{\bar{\alpha}_t} x_0 + \sqrt{1 - \bar{\alpha}_t}\, \epsilon$ | Compute noisy input |
| | 5 | $L = \|\epsilon - \epsilon_\theta(x_t, t)\|^2$ | Compute DDPM Loss |
| | 6 | $\theta \leftarrow \theta - \eta \nabla_\theta L$ | Update model parameters |
| Physics-Guided Stage | 7 | $(x_0, c) \sim q(x_0, c)$ | Layout mask + target spectrum |
| | 8 | $t \sim$ 75% from [0,20], 25% from [21,100] | Biased timestep sampling |
| | 9 | $\epsilon \sim \mathcal{N}(0, I)$ | Gaussian noise |
| | 10 | $x_t = \sqrt{\bar{\alpha}_t} x_0 + \sqrt{1 - \bar{\alpha}_t}\, \epsilon$ | Forward diffusion |
| | 11 | $\hat{\epsilon} = \epsilon_\theta(x_t, t, \text{embed}(c))$ | U-Net noise prediction |
| | 12 | $\hat{x}_0 = \dfrac{x_t - \sqrt{1 - \bar{\alpha}_t}\, \hat{\epsilon}}{\sqrt{\bar{\alpha}_t}}$ | Denoised layout |
| | 13 | $L_{\text{Maxwell}}(E)$ $= \|\nabla \times (\mu_0^{-1} \nabla \times E) - \omega^2 \varepsilon_r E\|$, where $E = \hat{x}_0$ | Maxwell's loss |
| | 14 | $\theta \leftarrow \theta - \eta \nabla_\theta L_{\text{Maxwell}}(E)$ | Gradient descent on Maxwell's loss |
| | | **Until converged** | |

**Design Problem 1**

The first case study presented in this work focuses on the inverse design of gold nanostructures patterned on a silica substrate to achieve arbitrary, on-demand transmission spectra. The thickness of the gold layer is fixed at 40 nm, and the unit-cell periodicity is set to 320 nm in both the lateral dimensions, as illustrated in Figure 2a. The training dataset consists of 15,000 randomly generated gold patterns. Normally incident electromagnetic wave spanning wavelengths from 0.5 μm to 1.8 μm is used as the excitation source. In addition to the transmission response, the proper collection of electric field distribution plays a central role in the proposed framework. Specifically, the out-of-plane electric field component $E_z$ is collected at a plane positioned 5 nm below the bottom of the gold pattern, inside the glass. Because the electric field is sampled in close proximity to the metallic structure, its spatial distribution closely resembles the underlying geometry (as shown in figure 2b inset). At the same time, the electric field exhibits smoother boundaries and inherently satisfies Maxwell's equations, making it a physically meaningful and well-behaved intermediate representation.

    A mesh size of 10 nm is used for electric field sampling, resulting in a 33 × 33 electric field profile. Consequently, each data sample in the dataset comprises three components: a 33 × 33 electric field distribution, a 120-point transmission spectrum, and a 64 × 64 binary structure pattern. Two representative examples of the training dataset are shown in Figure 2c and 2d. The structures are intentionally designed to be complex and highly arbitrary in order to demonstrate the expressive capability and robustness of the proposed Maxwell-guided diffusion framework. The corresponding transmission spectra for these structures are also shown in Figure 2c and 2d, with the insets displaying the associated two-dimensional electric field distributions at the collection plane.

The electric field is recorded at the wavelength corresponding to the minimum transmittance value in order to maximize the electric field intensity. For each structure–field pair, it can be visually observed that the electric field pattern closely mirrors the original structure geometry (Figure 2a-2d). This strong correlation implies that once the electric field distribution is known, reconstructing the underlying structure becomes a significantly straightforward task. Leveraging this insight, the proposed MxDiffusion model first takes the 120-point transmission spectrum as input and generates the corresponding 33 × 33 electric field distribution as output. Compared to directly predicting complex structural geometries, predicting the electric field is a simpler and more stable learning task due to its smoother spatial variation and absence of sharp material boundaries. The uniformity of the electric field allows the diffusion model to focus on capturing the correct global pattern, rather than being hindered by abrupt boundary transitions of poorly optimized edge features. Importantly, Maxwell's loss enforces physical consistency in the boundary regions, which is critical for accurately reproducing the target optical response.

The proposed model is a conditional diffusion U-Net with the latest developed architectures. The core features of this model are four-level U-Net hierarchy, multi-head self-attention layer, coordinate channels, Classifier-Free Guidance (CFG) and Feature-wise Linear Modulation (FiLM).[40, 55-58] The model takes both timestep data and conditional spectral data and generates electric field patterns. Timestep information is encoded using sinusoidal time embeddings with an embedding dimension of 128. The spectral data are encoded using pure convolutional encoder followed by global average pooling and a Multi-Layer Perceptron (MLP) projection. The time and spectral embeddings are concatenated and projected into a unified conditioning vector, which is injected into every residual block using Feature-wise Linear Modulation (FiLM). This mechanism allows the network to modulate feature activations dynamically based on both the diffusion timestep and the target spectrum, significantly improving conditional expressiveness. Detailed model architecture is available in Supporting Information.

Figure 3 demonstrates the effectiveness of the proposed MxDiffusion framework through a direct comparison with a conventional data-driven diffusion model that does not incorporate Maxwell's loss. To ensure a fair evaluation, both models use identical network architectures and have the same complexity. Transmission spectra are selected from the held-out validation set for the comparison task. These spectra are first used as targets for the MxDiffusion framework to generate the corresponding 33 × 33 electric-field distributions. The predicted electric fields, together with the target spectra, are then used as conditional inputs to the second diffusion model to generate the final 64 × 64 structural designs. The generated structures are subsequently simulated using Lumerical FDTD to evaluate whether their transmission responses match the target spectra. Three randomly selected validation examples are shown for demonstration. In Figure 3a-3c, the target spectra and the simulated spectra of the generated structures obtained using MxDiffusion framework are plotted together, along with the Mean Absolute Error (MAE) between the target and obtained spectra. For comparison, Figure 3e-3g presents the corresponding results obtained using the data-driven diffusion model for the same target spectra.

The results indicate that MxDiffusion framework consistently outperforms the data-only diffusion model, particularly in the target resonance regions. While both models tend to generate structurally similar patterns for a given target spectrum, the MxDiffusion framework exhibits superior boundary refinement capability, which is critical for accurately satisfying resonance conditions. As shown in Figs. 3a-3c and 3e-3g, the spectral performance of the structures generated by MxDiffusion closely match the target spectra, especially near the resonance wavelengths. In contrast, although the data-driven model produces geometrically similar structures, its spectral responses deviate more significantly from the targets. Figures 3d and 3h present the performance of MxDiffusion and the data-driven model, respectively, evaluated across the entire validation set. The majority of samples generated by MxDiffusion exhibit low mean absolute error (MAE), whereas the structures produced by the solely data-driven model consistently show significantly higher errors. These validation target spectra follow a distribution similar to that of the training data, under which MxDiffusion outperforms the solely data-driven model.

The performance of the two models is next evaluated on completely unseen target spectra that lies outside the training distribution. Figure 4 presents the results for several filter responses, along with the corresponding generated structural patterns shown as insets. Figures 4a and 4e correspond to an ideal long-pass filter target. After sampling both models, the MxDiffusion framework accurately captures the cutoff frequency of the target spectrum, with only minor deviations in the transmission magnitude within the passband and stopband regions. In contrast, the data-driven diffusion model exhibits substantially poorer agreement with the target response. For the short-pass filter design, as shown in Figures 4b and 4f present, MxDiffusion again demonstrates superior performance. Although both models show comparable behavior in the passband, MxDiffusion achieves significantly improved suppression in the stopband region. In the notch filter design shown in figures 4c and 4g, MxDiffusion accurately reproduces the target resonance wavelength, whereas the data-driven model fails to align the resonance correctly. Similarly, for the band-pass filter in figures 4d and 4h, MxDiffusion consistently outperforms the baseline model in matching the desired spectral response. These results highlight the strong generalization capability of MxDiffusion beyond the training distribution. In the following design problem, a more centrally distributed dataset is employed to further demonstrate the ability of the MxDiffusion framework to generate entirely new and previously unseen structural patterns.

**Design Problem 2**

This design problem focuses on the realization of a highly tunable bandpass filter in the infrared spectral range using the phase-change material Ge–Sb–Se–Te (GSST).[59, 60] As illustrated in Figure 5a, GSST can be reversibly switched between amorphous and crystalline phases, leading to pronounced changes in its optical properties and, consequently, the device transmission response. The device geometry is shown in Figure 5b. A 40 nm thick gold nanostructure is patterned on top of a Silicon Nitride (SiN) substrate, and the entire structure is subsequently capped with a 300 nm GSST layer and the unit-cell periodicity is set to 900 nm in both the lateral dimensions. An electromagnetic wave is incident from the top of the structure, and the transmission spectrum is recorded at the bottom. The electric field is sampled at a plane located 1 nm below the interface between the gold pattern and the SiN substrate, inside the substrate. As in the previous case studies, each data sample consists of three components: the structural pattern image, the corresponding electric-field distribution, and the transmission spectrum.

In this study, the dataset is restricted to structures that exhibit bandpass filtering behavior. The dataset contains a total of 24,000 unique x-y symmetric samples. Both the amorphous and crystalline phases of GSST are simulated for each structure, and the transmission spectra for both phases are recorded. A key characteristic of the GSST-based design is that the resonance wavelength undergoes a red shift when the material transitions from the amorphous to the crystalline phase. The difference between the resonance wavelengths in the two phases is defined as the tunability of the filter. Figures 5c and 5d present two representative examples from the dataset. For each pattern, the transmission spectra corresponding to both GSST phases are plotted, while the associated gold pattern and electric-field distribution are shown in the insets. The resulting tunability for each design is also indicated, highlighting the strong phase-dependent spectral control enabled by the GSST-based platform.

The MxDiffusion framework is trained on the GSST-based bandpass filter dataset described above. In this case, each structure is associated with two transmission spectra, corresponding to the amorphous and crystalline phases of GSST. Since each spectrum consists of 120 sampling points, the total conditional input to the model is a 240-point target spectrum. As in the previous design problem, the first diffusion model is trained to predict the electric-field distribution from the target spectrum using a combination of the standard DDPM loss and the physics-based Maxwell's loss. The second diffusion model then generates the final structural pattern conditioned on the predicted electric field and the target spectra. Because the electric field closely resembles the underlying structure, the second-stage prediction is relatively straightforward, with the primary modeling burden handled by the first physics-driven stage.

The network architecture used for this problem is identical to that employed in the first case study, incorporating a four-level U-Net hierarchy, FiLM conditioning, a multi-head self-attention block, coordinate channels, and classifier-free guidance. The total number of trainable parameters is approximately 60 million. After training, the model is evaluated on a held-out validation set. Figures 5e and 5f present two representative validation examples, where the ground-truth structures and the generated structures are shown as insets. For each case, the MxDiffusion model generates the electric field from the target spectra, which is then used by the second diffusion model to produce the final structure. The generated structures are subsequently simulated using Lumerical FDTD, and the resulting transmission spectra are compared with the targets. The close agreement between the simulated and target spectra demonstrates the effectiveness of the proposed framework.

The primary objective of this design problem is to generate structures with highly tunable spectral behavior. To establish a baseline, the structure in the dataset with the highest tunability is first identified. The best tunable structure is shown in Figure 5g. This structure exhibits a tunability of 0.97 μm. A more ambitious target is then defined by manually shifting the amorphous-phase spectrum toward shorter wavelengths and the crystalline-phase spectrum toward longer wavelengths, resulting in a target tunability of 1.21 μm. Using this target, the MxDiffusion model generates a new structure, shown in the inset of Figure 5g, which achieves a tunability of 1.04 μm. Although this value does not fully reach the intentionally aggressive target, due to inherent physical constraints, it represents a substantial improvement over the best structure available in the dataset. Notably, this generated structure lies outside the original data distribution, yet the MxDiffusion model successfully predicts a design with superior performance. We also trained the model with data-driven diffusion model, but that model cannot go beyond the best structure from the dataset. These results highlight the ability of the MxDiffusion framework to extrapolate beyond existing designs and improve inverse-design performance beyond the limits of traditional data-driven approaches.

**Conclusion**

We have introduced a physics-guided diffusion-based inverse design framework that integrates Maxwell's equations as an intermediate physical validation constraint within the diffusion process. Incorporating Maxwell's law enables the model to explicitly enforce electromagnetic consistency during training, significantly improving its ability to refine structural boundary regions, which play a critical role in achieving precise resonance conditions and accurately satisfying target design specifications. The proposed framework is systematically evaluated against a conventional data-driven diffusion model using both validation samples drawn from the training distribution and completely unseen target spectra. In all cases, the MxDiffusion model consistently outperforms the solely data-driven baseline, with particularly pronounced improvements observed for out-of-training-distribution design objectives. By embedding physical insight directly into the learning process, the model is able to extrapolate beyond the training dataset and generate high-performance designs that are not present in the original data distribution. While the demonstrations in this work focus on two-dimensional pattern optimization, the proposed approach is naturally extensible to three-dimensional inverse structural design problems without conceptual modification, as three-dimensional electromagnetic field distributions are also governed by Maxwell's equations. Furthermore, this physics-guided diffusion strategy is broadly applicable to a wide range of photonic design tasks, including photonic crystals and multilayer metasurfaces. The design objective is not limited to spectral responses; any physically measurable electromagnetic target can be incorporated. Thus, the proposed framework provides a general and extensible solution for physics-aware inverse photonic design for a variety of optical and nanophotonic design goals.

**References:**

(1) Zhu, D.; Liu, Z.; Raju, L.; Kim, A. S.; Cai, W. Building multifunctional metasystems via algorithmic construction. *ACS Nano* **2021**, *15* (2), 2318-2326.

(2) Ma, W.; Liu, Z.; Kudyshev, Z. A.; Boltasseva, A.; Cai, W.; Liu, Y. Deep learning for the design of photonic structures. *Nature Photonics* **2021**, *15* (2), 77-90.

(3) So, S.; Badloe, T.; Noh, J.; Bravo-Abad, J.; Rho, J. Deep learning enabled inverse design in nanophotonics. *Nanophotonics* **2020**, *9* (5), 1041-1057.

(4) Chen, Y.; Montes McNeil, A.; Park, T.; Wilson, B. A.; Iyer, V.; Bezick, M.; Choi, J.-I.; Ojha, R.; Mahendran, P.; Singh, D. K. Machine-learning-assisted photonic device development: a multiscale approach from theory to characterization. *Nanophotonics* **2025**, *14* (23), 3761-3793.

(5) Liu, Z.; Zhu, D.; Raju, L.; Cai, W. Tackling photonic inverse design with machine learning. *Advanced Science* **2021**, *8* (5), 2002923.

(6) Jiang, J.; Chen, M.; Fan, J. A. Deep neural networks for the evaluation and design of photonic devices. *Nature Reviews Materials* **2021**, *6* (8), 679-700.

(7) Park, T.; Mondal, S.; Cai, W. Interfacing nanophotonics with deep neural networks: AI for photonic design and photonic implementation of AI. *Laser & Photonics Reviews* **2025**, *19* (8), 2401520.

(8) Marzban, R.; Adibi, A.; Pestourie, R. Inverse design in nanophotonics via representation learning. *Advanced Optical Materials* **2026**, *14* (1), e02062.

(9) Li, Z.; Pestourie, R.; Lin, Z.; Johnson, S. G.; Capasso, F. Empowering metasurfaces with inverse design: principles and applications. *ACS Photonics* **2022**, *9* (7), 2178-2192.

(10) Khaireh-Walieh, A.; Langevin, D.; Bennet, P.; Teytaud, O.; Moreau, A.; Wiecha, P. R. A newcomer's guide to deep learning for inverse design in nano-photonics. *Nanophotonics* **2023**, *12* (24), 4387-4414.

(11) Kim, J.; Kim, J.-Y.; Kim, J.; Hyeong, Y.; Neseli, B.; You, J.-B.; Shim, J.; Shin, J.; Park, H.-H.; Kurt, H. Inverse design of nanophotonic devices enabled by optimization algorithms and deep learning: recent achievements and future prospects. *Nanophotonics* **2025**, *14* (2), 121-151.

(12) Malkiel, I.; Mrejen, M.; Nagler, A.; Arieli, U.; Wolf, L.; Suchowski, H. Plasmonic nanostructure design and characterization via deep learning. *Light: Science & Applications* **2018**, *7* (1), 60.

(13) Liu, D.; Tan, Y.; Khoram, E.; Yu, Z. Training deep neural networks for the inverse design of nanophotonic structures. *ACS Photonics* **2018**, *5* (4), 1365-1369.
(14) Molesky, S.; Lin, Z.; Piggott, A. Y.; Jin, W.; Vucković, J.; Rodriguez, A. W. Inverse design in nanophotonics. *Nature Photonics* **2018**, *12* (11), 659-670.
(15) An, S.; Zheng, B.; Shalaginov, M. Y.; Tang, H.; Li, H.; Zhou, L.; Ding, J.; Agarwal, A. M.; Rivero-Baleine, C.; Kang, M. Deep learning modeling approach for metasurfaces with high degrees of freedom. *Optics Express* **2020**, *28* (21), 31932-31942.
(16) Zhang, H.; Kang, L.; Campbell, S. D.; Young, J. T.; Werner, D. H. Data driven approaches in nanophotonics: A review of AI-enabled metadevices. *Nanoscale* **2025**, 17, 23788-23803.
(17) Raju, L.; Lee, K.-T.; Liu, Z.; Zhu, D.; Zhu, M.; Poutrina, E.; Urbas, A.; Cai, W. Maximized frequency doubling through the inverse design of nonlinear metamaterials. *ACS Nano* **2022**, *16* (3), 3926-3933.
(18) Adibnia, E.; Mansouri-Birjandi, M. A.; Ghadrdan, M.; Jafari, P. A deep learning method for empirical spectral prediction and inverse design of all-optical nonlinear plasmonic ring resonator switches. *Scientific Reports* **2024**, *14* (1), 5787.
(19) Ji, W.; Chang, J.; Xu, H.-X.; Gao, J. R.; Gröblacher, S.; Urbach, H. P.; Adam, A. J. Recent advances in metasurface design and quantum optics applications with machine learning, physics-informed neural networks, and topology optimization methods. *Light: Science & Applications* **2023**, *12* (1), 169.
(20) Kudyshev, Z. A.; Shalaev, V. M.; Boltasseva, A. Machine learning for integrated quantum photonics. *Acs Photonics* **2020**, *8* (1), 34-46.
(21) Liu, G.-X.; Liu, J.-F.; Zhou, W.-J.; Li, L.-Y.; You, C.-L.; Qiu, C.-W.; Wu, L. Inverse design in quantum nanophotonics: combining local-density-of-states and deep learning. *Nanophotonics* **2023**, *12* (11), 1943-1955.
(22) Yin, S.; Cao, R.; Liang, M.; Shen, C.; Zhou, H.; Zhang, O.; Yang, C. Can deep neural networks work with amplitude and phase input of defocused images? *Optics Express* **2024**, *32* (14), 25036-25045.
(23) Li, L.; Ruan, H.; Liu, C.; Li, Y.; Shuang, Y.; Alù, A.; Qiu, C.-W.; Cui, T. J. Machine-learning reprogrammable metasurface imager. *Nature Communications* **2019**, *10* (1), 1082.
(24) Wu, Q.; Xu, Y.; Zhao, J.; Liu, Y.; Liu, Z. Localized plasmonic structured illumination microscopy using hybrid inverse design. *Nano Letters* **2024**, *24* (37), 11581-11589.

(25) Christensen, T.; Loh, C.; Picek, S.; Jakobović, D.; Jing, L.; Fisher, S.; Ceperic, V.; Joannopoulos, J. D.; Soljačić, M. Predictive and generative machine learning models for photonic crystals. *Nanophotonics* **2020**, *9* (13), 4183-4192.

(26) Zhelyeznyakov, M. V.; Brunton, S.; Majumdar, A. Deep learning to accelerate scatterer-to-field mapping for inverse design of dielectric metasurfaces. *ACS Photonics* **2021**, *8* (2), 481-488.

(27) Qian, C.; Zheng, B.; Shen, Y.; Jing, L.; Li, E.; Shen, L.; Chen, H. Deep-learning-enabled self-adaptive microwave cloak without human intervention. *Nature Photonics* **2020**, *14* (6), 383-390.

(28) Wiecha, P. R.; Muskens, O. L. Deep learning meets nanophotonics: a generalized accurate predictor for near fields and far fields of arbitrary 3D nanostructures. *Nano Letters* **2019**, *20* (1), 329-338.

(29) Ma, W.; Cheng, F.; Xu, Y.; Wen, Q.; Liu, Y. Probabilistic representation and inverse design of metamaterials based on a deep generative model with semi-supervised learning strategy. *Advanced Materials* **2019**, *31* (35), 1901111.

(30) Kudyshev, Z. A.; Kildishev, A. V.; Shalaev, V. M.; Boltasseva, A. Machine learning–assisted global optimization of photonic devices. *Nanophotonics* **2020**, *10* (1), 371-383.

(31) Liu, Z.; Zhu, D.; Rodrigues, S. P.; Lee, K.-T.; Cai, W. Generative model for the inverse design of metasurfaces. *Nano Letters* **2018**, *18* (10), 6570-6576.

(32) An, S.; Zheng, B.; Tang, H.; Shalaginov, M. Y.; Zhou, L.; Li, H.; Kang, M.; Richardson, K. A.; Gu, T.; Hu, J. Multifunctional metasurface design with a generative adversarial network. *Advanced Optical Materials* **2021**, *9* (5), 2001433.

(33) Kim, W.; Kim, S.; Lee, M.; Seok, J. Inverse design of nanophotonic devices using generative adversarial networks. *Engineering Applications of Artificial Intelligence* **2022**, *115*, 105259.

(34) Li, Z.; Zhou, Z.; Qiu, C.; Chen, Y.; Liang, B.; Wang, Y.; Liang, L.; Lei, Y.; Song, Y.; Jia, P. The intelligent design of silicon photonic devices. *Advanced Optical Materials* **2024**, *12* (7), 2301337.

(35) Jiang, J.; Fan, J. A. Global optimization of dielectric metasurfaces using a physics-driven neural network. *Nano Letters* **2019**, *19* (8), 5366-5372.

(36) Chen, M.; Lupoiu, R.; Mao, C.; Huang, D.-H.; Jiang, J.; Lalanne, P.; Fan, J. A. High speed simulation and freeform optimization of nanophotonic devices with physics-augmented deep learning. *ACS Photonics* **2022**, *9* (9), 3110-3123.

(37) Lynch, S.; LaMountain, J.; Fan, B.; Bu, J.; Raju, A.; Wasserman, D.; Karpatne, A.; Podolskiy, V. A. Physics-guided hierarchical neural networks for Maxwell's equations in plasmonic metamaterials. *ACS Photonics* **2025**, *12* (8), 4279-4288.

(38) Sarkar, S.; Ji, A.; Jermain, Z.; Lipton, R.; Brongersma, M.; Dayal, K.; Noh, H. Y. Physics-informed machine learning for inverse design of optical metamaterials. *Advanced Photonics Research* **2023**, *4* (12), 2300158.

(39) Shao, G.; Zhou, T.; Yan, T.; Guo, Y.; Zhao, Y.; Huang, R.; Fang, L. Reliable, efficient, and scalable photonic inverse design empowered by physics-inspired deep learning. *Nanophotonics* **2025**, *14* (16), 2799-2810.

(40) Liu, R.; Lehman, J.; Molino, P.; Petroski Such, F.; Frank, E.; Sergeev, A.; Yosinski, J. An intriguing failing of convolutional neural networks and the coordconv solution, in *Proceedings of the 32nd Conference on Neural Information Processing Systems (NIPS)*, Montréal, QC, Canada, 3–8 December **2018**; pp. 9628–9639.

(41) Zhang, Z.; Yang, C.; Qin, Y.; Feng, H.; Feng, J.; Li, H. Diffusion probabilistic model based accurate and high-degree-of-freedom metasurface inverse design. *Nanophotonics* **2023**, *12* (20), 3871-3881.

(42) Hen, L.; Yosef, E.; Raviv, D.; Giryes, R.; Scheuer, J. Inverse design of diffractive metasurfaces using diffusion models. *ACS Photonics* **2025**, *13* (1), 38-46.

(43) Kim, J.; Neseli, B.; Yoon, J.; Kim, J. Y.; Hong, S.; Park, H. H.; Kurt, H. Semi-supervised learning leveraging denoising diffusion probabilistic models for the characterization of nanophotonic devices. *Laser & Photonics Reviews* **2024**, *18* (10), 2300998.

(44) Niu, C.; Phaneuf, M.; Mojabi, P. A diffusion model for multi-layered metasurface unit cell synthesis. *IEEE Open Journal of Antennas and Propagation* **2023**, *4*, 654-666.

(45) Zhang, Z.; Yang, C.; Qin, Y.; Zheng, Z.; Feng, J.; Li, H. Addressing high-performance data sparsity in metasurface inverse design using multi-objective optimization and diffusion probabilistic models. *Optics Express* **2024**, *32* (23), 40869-40885.

(46) Chakravarthula, P.; Sun, J.; Li, X.; Lei, C.; Chou, G.; Bijelic, M.; Froesch, J.; Majumdar, A.; Heide, F. Thin on-sensor nanophotonic array cameras. *ACM Transactions on Graphics (TOG)* **2023**, *42* (6), 1-18.

(47) Zhu, L.; Hua, W.; Lv, C.; Liu, Y. Rapid inverse design of high degree of freedom meta-atoms based on the image-parameter diffusion model. *Journal of Lightwave Technology* **2024**, *42* (15), 5269-5278.

(48) Sun, J.; Chen, X.; Wang, X.; Zhu, D.; Zhou, X. Photonic modes prediction via multi-modal diffusion model. *Machine Learning: Science and Technology* **2024**, *5* (3), 035069.

(49) Chen, Y.; Bezick, M.; Wilson, B.; Kildishev, A.; Shalaev, V.; Boltasseva, A. Physics-informed latent diffusion model for high-efficiency metasurface optimization; *Proc. SPIE, Photonic Computing: From Materials and Devices to Systems and Applications II*, San Diego, California, USA, 18 September **2025**, PC135810S.

(50) Seo, D.; Um, S.; Lee, S.; Ye, J. C.; Chung, H. Physics-guided and fabrication-aware inverse design of photonic devices using diffusion models. *ACS Photonics* **2025**, 13 (2), 363–372.

(51) Ho, J.; Jain, A.; Abbeel, P. Denoising diffusion probabilistic models. *Advances in Neural Information Processing Systems* **2020**, *33*, 6840-6851.

(52) Song, Y.; Sohl-Dickstein, J.; Kingma, D. P.; Kumar, A.; Ermon, S.; Poole, B. Score-based generative modeling through stochastic differential equations. *arXiv preprint arXiv:2011.13456* **2020**.

(53) Sohl-Dickstein, J.; Weiss, E.; Maheswaranathan, N.; Ganguli, S. Deep unsupervised learning using nonequilibrium thermodynamics. In *Proceedings of the 32nd International Conference on Machine Learning*, Lille, France, July 6 - 11, **2015**, PMLR, vol. 37, pp 2256-2265.

(54) Nichol, A. Q.; Dhariwal, P. Improved denoising diffusion probabilistic models. In *Proceedings of the 38th International Conference on Machine Learning*, Virtual Event, 18-24 July**, 2021**, PMLR, vol 139, pp 8162-8171.

(55) Ronneberger, O.; Fischer, P.; Brox, T. U-net: Convolutional networks for biomedical image segmentation. *In Proceedings of the 18th International Conference on Medical Image Computing and Computer-Assisted Intervention*, Munich, Germany, October 5-9, **2015**, Springer, pp 234-241.

(56) Vaswani, A.; Shazeer, N.; Parmar, N.; Uszkoreit, J.; Jones, L.; Gomez, A. N.; Kaiser, Ł.; Polosukhin, I. Attention is all you need. *In Proceedings of the 31st International Conference on Neural Information Processing Systems (NIPS),* Long Beach, CA, USA, December 4–9, **2017**, MIT Press, pp. 5998–6008.

(57) Ho, J.; Salimans, T. Classifier-free diffusion guidance. *arXiv preprint arXiv:2207.12598* **2022**.

(58) Perez, E.; Strub, F.; De Vries, H.; Dumoulin, V.; Courville, A. Film: Visual reasoning with a general conditioning layer. In *Proceedings of the 32nd AAAI conference on artificial intelligence*, February 2–7, New Orleans, Louisiana, USA, **2018**, vol. 32. pp 3942-3951

(59) Sahoo, D.; Naik, R. GSST phase change materials and its utilization in optoelectronic devices: A review. *Materials Research Bulletin* **2022**, *148*, 111679.

(60) Miscuglio, M.; Meng, J.; Yesiliurt, O.; Zhang, Y.; Prokopeva, L. J.; Mehrabian, A.; Hu, J.; Kildishev, A. V.; Sorger, V. J. Artificial synapse with mnemonic functionality using GSST-based photonic integrated memory. In *2020 international applied computational electromagnetics society symposium (ACES)*, Monterey, CA, USA, July 27–31, **2020**; IEEE: pp 1-3.

**Figures and Figure Captions**

**Figure 1**

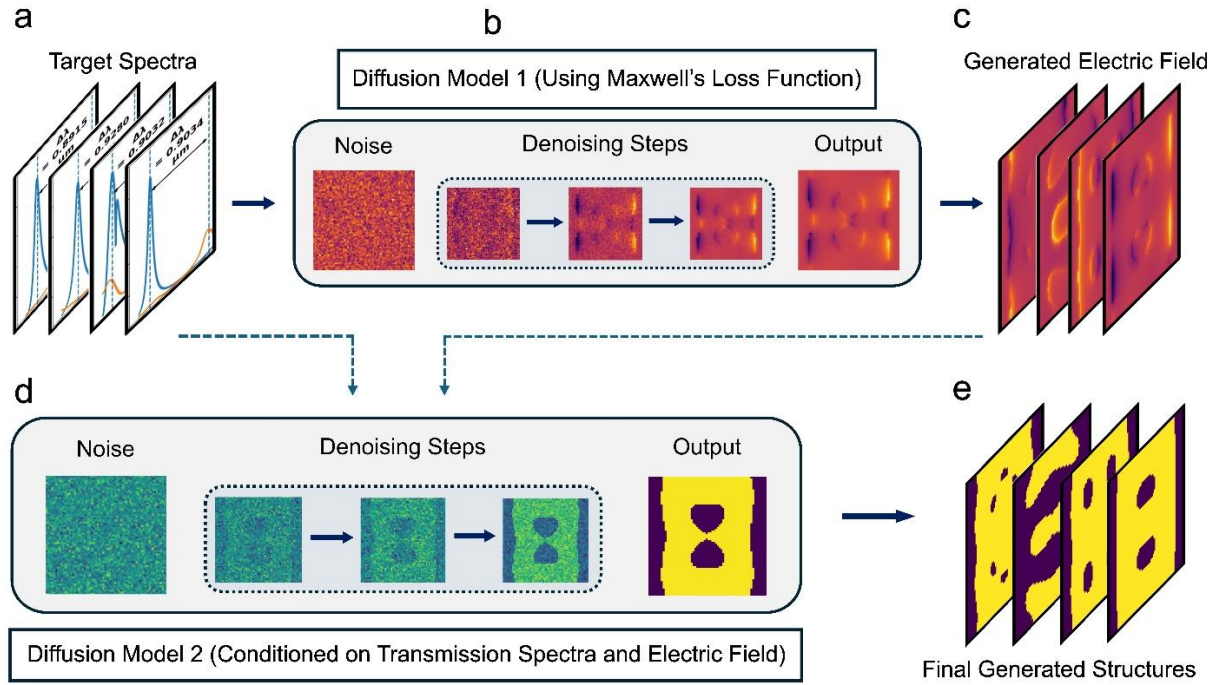

**Figure 1: MxDiffusion model architecture.** (a) Target transmission spectra. The objective of the model architecture is to generate 2D structure patterns that can satisfy these target spectra. (b) Diffusion model 1. This model uses Maxwell's loss to predict the electric field from the spectral data input. (c) Generated electric field pattern from the first diffusion model. (d) Diffusion model 2. This model uses both the initial target spectral data and the generated electric field as conditional inputs to produce the final structure. (e) Final generated patterns.

**Figure 2**

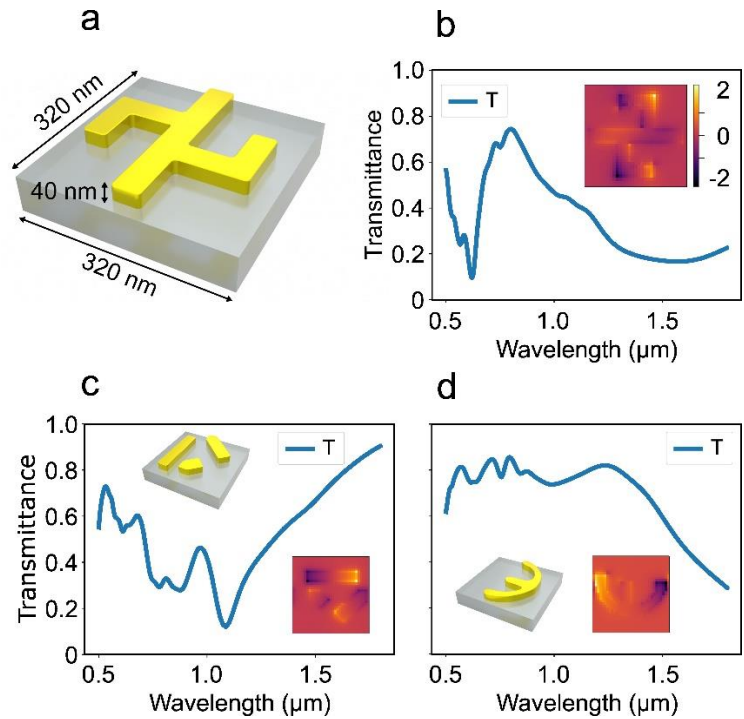

**Figure 2: Dataset generation process for design problem 1.** (a) Unit cell consisting of the gold pattern on a silica substrate. (b) Corresponding transmission spectra for the structure in part (a). The inset represents the electric field collected just below the gold-silica interface inside the substrate. (c-d) Simulated spectra of two representative gold patterns from the training dataset. The gold patterns are shown in the insets, highlighting the complexity and diversity of the structural designs. The associated two-dimensional electric-field distributions are also shown as insets.

Figure 3

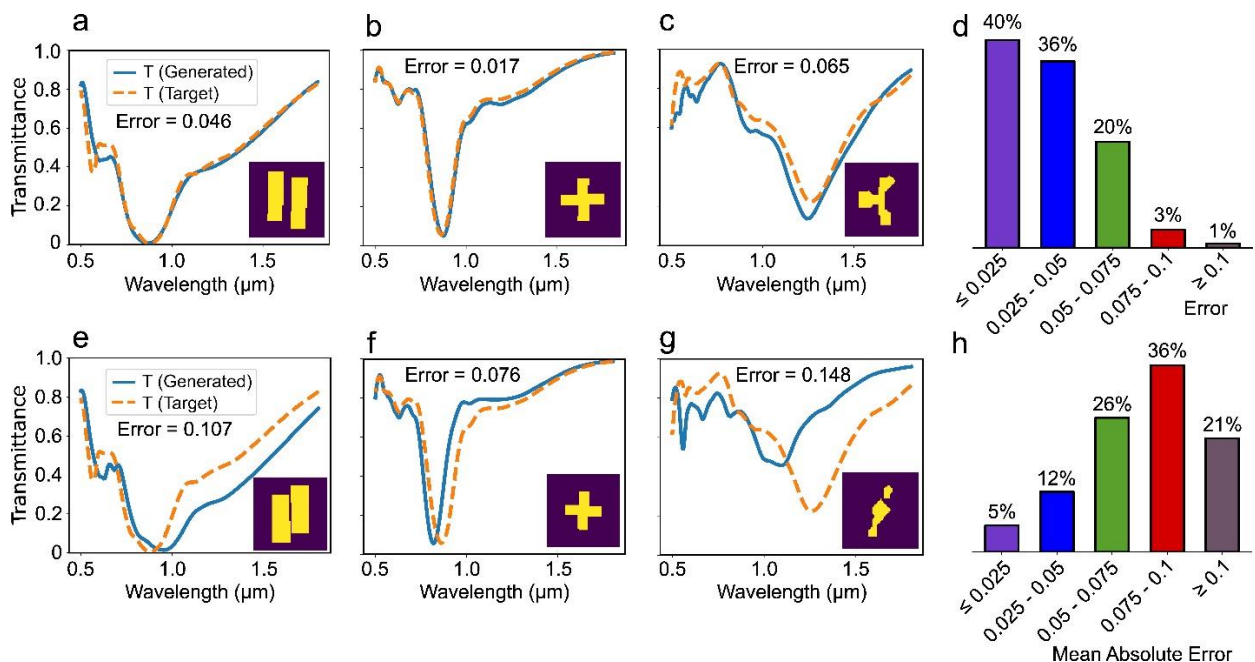

**Figure 3: Performance comparison between the MxDiffusion model and solely data driven diffusion model on validation set.** (a-c) Prediction performance of the MxDiffusion framework. Several target transmission spectra are randomly selected from the validation set and used as conditional inputs to the MxDiffusion architecture to produce the final structural designs (shown as insets). The generated structures are subsequently simulated to obtain their transmission spectra, which closely match the target spectra for most samples. (d) Error diagram for MxDiffusion model on validation set of 1464 samples. Most samples have very low errors. (e-g) Performance of the solely data-driven diffusion model, where the same target spectra from (a-c) are used as inputs. (h) Error diagram for the solely data-driven diffusion model. In comparison, MxDiffusion consistently achieves significantly better agreement with the target spectra than the data-driven approach.

**Figure 4**

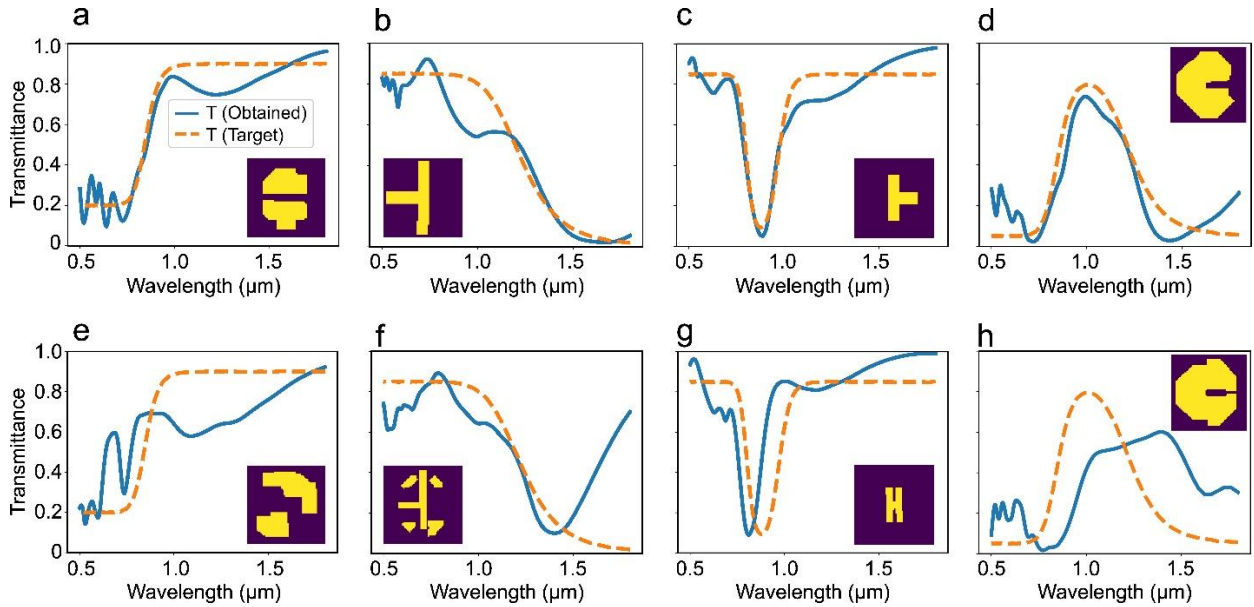

**Figure 4: Performance comparison between the MxDiffusion model and solely data driven diffusion model on out-of-training-distribution target spectra**. (a-d) Target transmission spectra and simulated transmission spectra of structures generated by the MxDiffusion framework for (a) long-pass, (b) short-pass, (c) notch, and (d) bandpass filter responses. The generated patterns are shown as insets. The transmission spectra of the generated patterns accurately match the cutoff or resonance frequencies in all cases, while minor deviations in passband and stopband levels arise due to inherent physical limitations. (e-h) Corresponding results obtained using the solely data-driven diffusion model for the same target spectra, exhibiting significantly larger discrepancies from the desired responses.

**Figure 5**

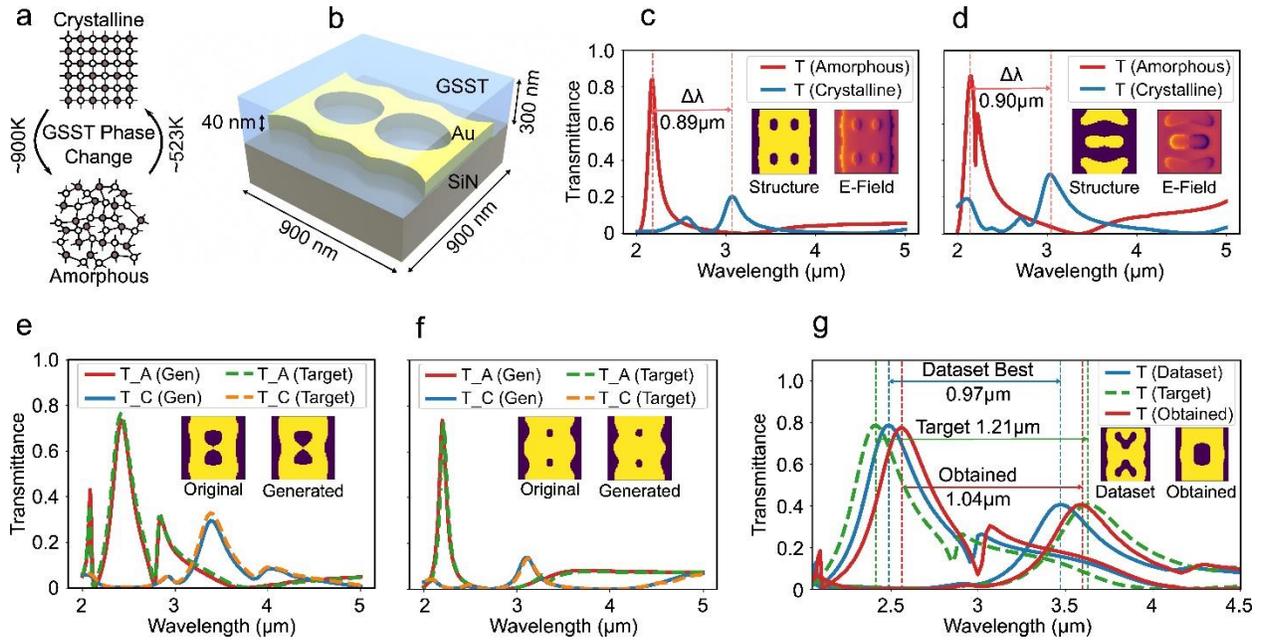

**Figure 5: Schematic of the GSST-based structure and the performance of MxDiffusion model on this dataset.** (a) Illustration of the phase transition of GSST material from the amorphous to the crystalline state. (b) Unit cell structure of the GSST-based device. (c-d) Representative sample structures (inset) from the dataset, where the transmission spectra for both amorphous and crystalline phases are plotted along with the resulting tunability. The corresponding electric-field distributions are also shown as insets. (e-f) Performance of the trained MxDiffusion framework on two randomly selected validation samples, showing close agreement between the target and generated transmission spectra; the original and generated structural patterns are displayed as insets. (g) Design performance of MxDiffusion for an out-of-training-distribution, highly ambitious tunability target. Here same color is used for the two transmission curves of the two phases for a particular structure. We start from the best structure in the dataset with a spectral tunability of 0.97 μm, then the amorphous and crystalline spectra are shifted to left and right respectively defining an aggressive target. This target spectrum is used in the MxDiffusion framework, resulting in a generated structure with a spectral tunability of 1.04 μm, an improvement of 70 nm over the best dataset sample.